\pdfoutput=1
\documentclass{vgtc}                          

\ifpdf
  \pdfoutput=1\relax                   
  \usepackage{graphicx}                
  \DeclareGraphicsExtensions{.pdf,.png,.jpg,.jpeg} 
\else
  \usepackage{graphicx}                
  \DeclareGraphicsExtensions{.eps}     
\fi%

\graphicspath{{figures/}{pictures/}{images/}{./}} 

\usepackage{microtype}                 
\PassOptionsToPackage{warn}{textcomp}  
\usepackage{textcomp}                  
\usepackage{mathptmx}                  
\usepackage{times}                     
\usepackage{cite}                      
\usepackage{tabu}                      
\usepackage{booktabs}                  

\onlineid{0}

\vgtccategory{Research}

\vgtcinsertpkg

\title{Data Visualization Practitioners' Perspectives on Chartjunk}

\author{Paul Parsons\thanks{e-mail: parsonsp@purdue.edu} %
\and Prakash Shukla\thanks{e-mail: shukla37@purdue.edu}}
\affiliation{\scriptsize Purdue University}

\abstract{Chartjunk is a popular yet contentious topic. Previous studies have shown that extreme minimalism is not always best, and that visual embellishments can be useful depending on the context. While more knowledge is being developed regarding the effects of embellishments on users, less attention has been given to the perspectives of practitioners regarding how they design with embellishments. We conducted semi-structured interviews with 20 data visualization practitioners, investigating how they understand chartjunk and the factors that influence how and when they make use of embellishments. Our investigation uncovers a broad and pluralistic understanding of chartjunk among practitioners, and foregrounds a variety of personal and situated factors that influence the use of chartjunk beyond context. We highlight the personal nature of design practice, and discuss the need for more practice-led research to better understand the ways in which concepts like chartjunk are interpreted and used by practitioners.  %
} 

\CCScatlist{
  \CCScatTwelve{Human-centered computing}{Visu\-al\-iza\-tion}{}{};
}

\begin{document}


\firstsection{Introduction}

\maketitle


Chartjunk has been a popular yet contentious concept for many decades. Since its inception there has been considerable debate about the value of visual embellishments, whether they have any proper role in visualization design, and under what conditions they may be appropriate to use. In the years after the term was coined by Tufte \cite{tufte_visual_1983}, only a small number of research studies investigated chartjunk (e.g., \cite{tractinsky_chartjunk_1999,blasio_comparison_2002,inbar_minimalism_2007}). In contrast, the past decade has seen a renewed interest in the InfoVis research community \cite{bateman_useful_2010,borgo_empirical_2012,borkin_what_2013,borkin_beyond_2016,li_is_2014,quispel_would_2014,haroz_isotype_2015,hill_minimalism_2017,pena_memorability_2020,Correll_Gleicher_2014}. The vast majority of this research has focused on how users perceive chartjunk, largely with respect to comprehension, memorability, and performance measures such as speed and accuracy. While these studies have contributed valuable knowledge about the effects of embellishments on users, there has been little attempt to study embellishments from the perspective of visualization designers. Studying the effects of a phenomenon on users provides at best a partial understanding of its role in design, as there are many factors that influence the use of concepts and principles in real-world settings.

Designers have ways of knowing and thinking that are distinct from those of scientists and researchers---famously referred to by Nigel Cross as \textit{designerly ways of knowing} \cite{cross_designerly_1982}. Studies of design practice in other emergent design fields, such as interaction design and instructional design, indicate that designers confront the complexity of real-world situations in ways that are quite different from those of researchers \cite{goodman_understanding_2011,Stolterman2008}. While a focus on design practice is gaining traction in InfoVis \cite{bigelow2016iterating,mendez2017bottom,hoffswell2020techniques,walny2019data,parsons_what_2020,Parsons_judgement_2020} there is still not much investigation into how designers generate and use knowledge in practice. As chartjunk is one of the most well known and used concepts by practitioners \cite{parsons_what_2020}, it presents a good opportunity to pursue such an investigation.

We conducted interviews with 20 DataVis practitioners, in which they were asked about their perspectives on the concept of chartjunk and the ways in which it influences their design practice. In doing so, we surface the concerns of practitioners, highlighting issues surrounding the \textit{creation of embellishments} and not only their \textit{effects on users}. We contribute a practice-led perspective to the extant literature, highlighting the personal and situated aspects of design practice that are not considered in typical user studies. 
 
\section{Chartjunk}
Tufte is credited with coining the term chartjunk in his 1983 book The Visual Display of Quantitative Information \cite{tufte_visual_1983}. He defined it as ``ink that does not tell the viewer anything new'' and ``non-data-ink or redundant data-ink''. Tufte defined \textit{data-ink} as ``the non-erasable core of a graphic, the non-redundant ink arranged in response to variation in the numbers represented'', and the \textit{data--ink ratio} as the ratio of the data-ink over the total ink used in a graphic \cite{tufte_visual_1983}. Tufte's ``theory of data graphics'' proposed maximizing the data-ink ratio, which is another way of saying to reduce chartjunk---i.e., avoid non-data-ink. Despite Tufte's extreme minimalism, he did admit a place for considering ``complexity, structure, density, and even beauty'' in the design of graphics, seemingly opening opportunities where non-data-ink may be acceptable, although no clarity was provided on how and what this might look like.  

Despite Tufte's highly influential work, many information and graphic designers---most notably Nigel Holmes---continued to create graphics with embellishments, leading to a fierce debate among two camps with opposing views \cite{few_chartjunk_2011}. Yet, as noted by Few \cite{few_chartjunk_2011}, little of the debate was dispassionate and evidence-based. 



\subsection{Prior Research}
Studies have investigated effects of chartjunk on users (often using more neutral labels such as embellishment or decoration), highlighting some benefits of embellishments \cite{bateman_useful_2010,gillan_minimalism_2009,borkin_beyond_2016,borkin_what_2013,haroz_isotype_2015,inbar_minimalism_2007}. For instance, Bateman et. al \cite{bateman_useful_2010} found embellished charts to improve recall (though there have been concerns about their methodology \cite{few_chartjunk_2011}). Borkin et. al \cite{borkin_what_2013,borkin_beyond_2016} have found that visualizations with scenes, people, and pictograms can be more memorable than simple charts. Haroz et al. \cite{haroz_isotype_2015} also investigated effects of pictographic representations on memory, speed, and engagement, finding that superfluous images can be distracting--but do not incur any significant user costs--and have benefits for working memory and engagement. 


From a different perspective, Hullman et. al \cite{hullman_benefitting_2011} have argued that adding desirable difficulties (e.g., embellishments) can augment comprehension and recall. Others have argued that going beyond minimalist chart design can improve engagement \cite{hung_assessing_2017,kostelnick2008visual}. Correll and Gleicher \cite{Correll_Gleicher_2014} have highlighted the reductionist view of the data-ink ratio as a guide for design, instead proposing less prescriptive guidelines that can be applied using a designer's best judgment.

One takeaway from these studies is that a balance between extreme minimalism and embellishments should be sought in a contextually appropriate manner. While this may seem like a simple guideline, its application in practice is not well understood. In this work we attempt to uncover some ways practitioners think about chartjunk and its use in real-world design settings.

\section{Method}
As part of a broader research effort to study DataVis design practice, we interviewed 20 DataVis practitioners and asked them about their design process, including how they understand and make use of specific concepts and methods. Recruiting was done via social media, the DataVis Society’s Slack workspace, the InfoVis email list, and our personal networks. To mitigate sampling bias, we also searched widely online for practicing professionals and agencies, ultimately contacting more than 200 individuals and more than 30 agencies.

Interviews were semi-structured and were conducted remotely via videoconferencing. For this paper, we selected one section of the transcripts that focused on design methods and principles, in which we asked participants: (1) whether they were familiar with the  concept of chartjunk or visual embellishment (we specifically included ``embellishment'' in an attempt to not frame the discussion too negatively); (2) if and how they relied on the idea of chartjunk/embellishment in their design work; (3) whether they had any opinions about chartjunk/embellishment; and (4) whether they were familiar with the history of the concept and its discourse in the academic or practitioner spaces. This section accounted for roughly 10 minutes near the middle of the 60-75 minutes of each interview. The transcripts were inductively coded by the two researchers, following standard processes for thematic analysis \cite{Braun_Clarke_2006}. We went through several rounds of open coding independently, regularly meeting with the goal of establishing agreement on the codes. We subsequently conducted several rounds of searching for and defining themes, eventually reaching consensus.

\begin{table}
\begin{tabular}{ p{0.05\columnwidth} | p{0.42\columnwidth} | p{0.09\columnwidth} | p{0.13\columnwidth} | p{0.06\columnwidth} }
\textbf{ID} & \textbf{Job Title} & \textbf{Exp. (yrs)} & \textbf{Highest Degree} & \textbf{G} \\ 
\hline
P1 & Sr. DataVis Designer & $>$10 & D & F \\
P2 & Data Storyteller & $>$10 & M & M \\
P3 & DataVis Engineer & 8-10 & D & M \\
P4 & DataVis/UX Designer & $>$10 & M & M \\
P5 & DataVis/UX Designer & 2-4 & M & F \\
P6 & DataVis Designer & 5-7 & M & F \\
P7 & DataVis Designer/Dev & $>$10 & M & M \\
P8 & DataVis Designer & 8-10 & B & M \\
P9 & Graphics Editor & 8-10 & B & F \\
P10 & DataVis Designer & 5-7 & B & F \\
P11 & Sr. UX Design Lead & 8-10 & D & M \\
P12 & DataVis Designer & 8-10 & M & M \\
P13 & DataVis Designer & 5-7 & B & M \\
P14 & Data Architect & $>$10 & M & M \\
P15 & Sr. UX/DataVis Designer & 5-7 & M & F \\
P16 & Data Communicator & 2-4 & M & M \\
P17 & DataVis Designer & 5-7 & M & F \\
P18 & DataVis Journalist/Designer & 5-7 & M & M \\
P19 & Sr. DataVis Dev & $>$10 & D & F \\
P20 & DataVis Designer & $>$10 & M & M \\
\hline
\end{tabular}
    \caption{Our 20 participants and their self-reported characteristics: job title, years of experience, highest degree (Bachelors-B, Masters-M and Doctoral-D) and gender. ~\label{tab:participants}}
\end{table}

\section{Findings}

\subsection{How Practitioners Understand Chartjunk}
\subsubsection{``Corrective'' Movement}
When asked about their perspectives on chartjunk, multiple participants mentioned there was somewhat of a corrective movement taking place, away from Tufte's extreme minimalism and toward an acceptance of more embellishment. For example, P15 stated ``\textit{I think that there's sort of this very minimalist, everything must be there for a reason, school of data visualization design. And I think hopefully that we've at least started to move past that}''. P18 similarly noted that ``\textit{Tufte was really strong in opposing chart junk. And I think now there's a movement going a little bit in the other direction saying that, well, adding icons to charts can make things clearer, for example. So yeah, it's bit like a pendulum, I think. And with Tufte it went in one direction and now it's going a little bit to back in the other direction.}'' P19, who has a PhD in visualization and is now a practitioner, commented on both the practitioner and researcher communities: ``\textit{Certainly I think it's been a necessary corrective to the way things were in the Vis community---both in the practitioner and I think in the research community---because the practitioners were so influenced by Tufte and because the research community was so influenced by the Cleveland and McGill idea of being just focused on the data.}''

\subsubsection{Conceptual and Terminological Issues}
Regarding the ways in which participants interpret and define chartjunk, we found several significant conceptual and terminological issues. For instance, many participants noted the obvious negative framing of the term, with P12 stating ``\textit{First of all, acknowledging that it's called junk---who argues for junk?}'' and P15 noting that ``\textit{It sounds like a negative connotation because it is literally calling anything extra trash.}''

There was little consensus on the definition of chartjunk, with a wide variety of interpretations. Aware of this, P16 stated ``\textit{I don't know if people have strong opinions about what is and isn't chart junk}'', and P6 noted ``\textit{I do think the definition is different [\ldots] I'm not 100 percent sure that everyone knows, like thinks the same thing when they think of chart junk.}'' P17 pointed to an ``infographic'' kind of interpretation, stating ``\textit{It might also be the definition for me personally---when I find chartjunk, that's more for me over these sort of infographic posters where there's the idea that there's a lot of graphic stuff.}'' P13 described confusion about whether chartjunk refers only to distortions or all ornamentation ``\textit{Now, if we're talking about distorting actual charts and the way that information is perceived, I definitely don't do that. But if there's some sort of ornamentation around the chart to draw your eye in or keep you engaged [\ldots] then I think that's a different purpose.}'' 

Beyond the breadth of interpretation regarding what visual elements are considered as chartjunk, multiple participants talked about animation and interactivity as potential forms of chartjunk. For instance,  P11 states ``\textit{that's [chartjunk] just another side of the data-ink-ratio conversation. But it goes beyond that, because I find people love to add little icons. You know, they're constantly adding little icons to buttons and user interactions and, you know, do they support understanding?}'' P13, when describing when embellishments might be desirable, gave an example focused on interaction ``\textit{if you moved your mouse along the screen and it, I dunno, changes colors, or when you click it radiates out [\ldots] maybe it's helpful.}'' P9, when describing coworkers who tend to desire chartjunk, described them in this way: ``\textit{they want stuff online that moves and hovers and does all this stuff, and we have to explain to them, well you have to hover---if the data isn't there, someone has to click on it to see the data [\ldots] chart junk is always there, man.}''

Finally, many participants noted the subjective nature of defining chartjunk. For instance, when describing the work of a well-known practitioner, P6 indicated that ``\textit{people wouldn't call it chartjunk, people would call it beautiful.}'' P9, when describing a previous project involving annotations on a series of charts, stated ``\textit{I wouldn't call that chart junk at that point. I would just call it a call out---annotations. }'' When asked about their view of chartjunk, P11 stated ``\textit{it depends on what you call chart junk. Like when I look at The New York Times data visualization, I don't think there's really any junk there. They do have things that are sometimes whimsical, or The Pudding does this too, it's always motivated.}''

\subsection{Factors Influencing the Use of Chartjunk}
\subsubsection{Balance and Context are Important}
When describing the use of embellishments, most participants described a need for ``balance''. For instance, according to P5, ``\textit{I try to strike a balance between making something look nice and kind of fun and cute versus just polluting it with a whole bunch of nonsense.}'' Similarly, P4 mentioned ``\textit{So of course there's this balance that you want to also provide this kind of context and help texts and labels and so on as well as possible, so that the users can use it}''. Many participants also referred to this notion of context, indicating that it is a significant factor. For example, P4 stated ``\textit{I'm really relativistic about this, that it depends on the case and the context [\ldots] I mean, I'm fine with data art or anything, but if somebody suggests that I should put that in some business dashboards, then I would kind of question that this is not the right place for it.}'' P20 similarly noted ``\textit{I have no problem with chart junk for certain audiences, certain applications, certain types of media---go for it.}'' P18 simply stated that ``\textit{the context determines how far you can go with it.}''

\subsubsection{More than Just Context}
Although context was noted as important, there were multiple issues surfaced beyond context. Participants described not using embellishments due to a lack of skill in creating them. For instance, P8 said ``\textit{I would say I may not be good at it because I just know how to write code and visually represent the data set. But if I have to come up with my own drawings or elements that are not part of the data, then maybe the challenge is a bit too big for me, to actually do those kinds of things.}'' P18 also noted that ``\textit{Maybe I should also mention that I don't know how to draw. [\ldots] I'm just really bad at making things around or outside the pure data visualization things. So I simply cannot produce chart junk, so to speak. I don't know how to draw. So for me it always focuses on the data.}'' P13 described making ``junk'' as a byproduct of attempting to create embellishments:  ``\textit{It's [embellishment] generally something that I try to avoid because I don't think that I do it well. So I think that you can embellish well if I, if I had tools to add texture to things, or if I had the illustration skills to add something that wasn't a distraction but felt like it could be an included part of something, then great. But my efforts with that tend to look more like junk than a part of the design. }''

Multiple participants also described their use of embellishments in relation to their own personal style, indicating influences beyond the impersonal context of the situation. For example, when describing a recent project, P8 described a design decision as being due to ``\textit{a style I personally liked.}'' When asked about the context in which embellishments might be used in their design process, P20 noted ``\textit{I lean slightly towards chart junk because I will always, especially in interactive projects, I believe in eye candy 100 percent. Like I'm going to animate that chart and bring it to life.}'' P18 described their style as ``\textit{a bit minimalistic. So data-ink ratio and chart junk are definitely things I try to think about when I design. But usually they are just in the back of my head---I don't have a list here that I have to check in each step just to make sure that I don't have too much chart junk in my work. It's a bit more already integrated into my style.}'' P17 described their ``stylistic'' approach to using embellishments:  ``\textit{I wouldn't say I'm a minimalist because the kinds of works that I do need to grab the reader's eye [\ldots] in terms of embellishments, I always try and keep it very close to the data. So instead of a color, I might make it a subtle gradient, I might give it a slight drop shadow to emphasize it [\ldots] Things like going beyond the default in a more stylistic way, that's kind of how I like to add my embellishments.}'' P19 described their style as not favoring embellishments: ``\textit{It's still probably not my style to design something that has a lot of embellishment to it. I'm generally a very `put the data first' kind of designer when it comes to working on datavis. I like having elements on the page, always going back to a data point in some way. I would rather introduce sort of visual variety by mapping something to multiple visual channels or having something done in the encoding to just sort of introduce a little extra color if it's necessary or to space things out some more.}''


Participants also often referenced a number of external constraints that influenced how and when they would use embellishments. For instance, multiple participants talked about using embellishments to maintain brand fidelity. When defining chartjunk, P14 stated ``\textit{chart junk speaks to two things [\ldots] composition, and [\ldots] this notion of brand fidelity, right?}'' later describing the goal of design as to ``\textit{build a good composition and build it with allegiance to your branding requirements.}'' P19, who works for a firm with a ``\textit{more stripped down}'' style, described embellishment as ``\textit{not something that comes up super often for us [\ldots] and you know, there's no real reason for that other than it's a branding thing.}'' P14 described their style as ``\textit{toward a clean minimalist approach}'', yet noting ``\textit{in order to maintain some sort of fidelity to a brand, there are times when throwing a bunch of teddy bears on the screen makes sense.}''

A group of participants also talked about having to consider the desires of clients when thinking about chartjunk. P20 mentioned ``\textit{ I know for a lot of my clients [\ldots] they may not want to do that [add embellishments] and sometimes you just want to show the data and have it clean and simple.}'' P4 described the challenge of pleasing clients who often want something that looks novel ``\textit{This can be a challenge with clients, because clients might expect some novel and special visualization. So then balancing that, can we now do something novel here, or can we just sell a well-thought bar chart to the users. So how to balance between these different needs.}''

\subsubsection{Cognitivist vs. Experiential Focus}
While the vast majority of participants indicated that embellishments had a place as long as they fit the context and other personal and situational goals, participants would regularly reveal underlying commitments to a particular design philosophy when justifying their views, often surfacing value-laden judgments about the proper ways in which visualizations should be designed and consumed. These typically manifested in terms of more cognitive or experiential concerns that were foregrounded as being important outcomes of users' interactions with their work. For instance, P9 described being generally against embellishment, stating that ``\textit{What does that [embellishment] help tell you? I'm not making these charts for someone to be like, oh, that's pretty cute. No, here's the locations, and then here's like the serious data---I'm not trying to be cutesy}'', revealing a rationalistic perspective on what gives value to a visualization and prioritizing ``seriousness'' over other potential metrics. Other participants foregrounded a cognitive focus while defending the use of embellishments, typically noting that they can enhance comprehension. For instance, when describing the value of embellishment, P5 noted ``\textit{if you could make your data pop, then it increases people's understanding of what they're looking at. And that's ultimately the idea behind data visualization---you want people to understand.}'' 

Other participants revealed more of a focus on experiential issues, often pointing to concepts such as engagement, beauty, fun, and enjoyment. P6 mentioned the famous ``monster'' chart by Nigel Holmes, stating that ``\textit{This chart I think is excellent [\ldots] it's awesome. I've seen many charts that with the same data, but if somebody wants to make a point for this chart, and they just show the data in a very simple line or bar chart, it's so boring}'', later stating that ``\textit{I'd say that making a chart more beautiful, with extra elements that don't distract from the actual methods, but are there more to attract potential readers, they're always helpful because it attracts readers or because it gives you a good feeling why you're reading this visualization.}'' P7, when describing work from a well-known practitioner with ``\textit{a lot of decorations going on}'' noted that ``\textit{it engages the audience}'' and that ``\textit{for those purposes they [embellishments]  make sense.}'' P10 also described using embellishments to convince viewers of the potential for DataVis, stating that ``\textit{I just want people to get interested, people that aren't in datavis for them to get interested in it, to see it as not only something extremely informative, but something that could potentially be beautiful. And I especially want students and younger kids, and especially kids that think that coding is boring or that it's like matrix-y, that this is something that could be beautiful and fun [\ldots] So I don't follow the purest data, like purest bar chart or anything like that.}''

\section{Discussion}

\subsection{Broad Interpretations of Chartjunk}
Our analysis shows that practitioners' understanding of chartjunk is very broad. While almost all participants knew that Tufte coined the term, and many knew about the data-ink ratio, there was no discernible consensus on the definition of chartjunk. Some participants described chartjunk as distortions of graphics, others as anything ``extra'', while others included animation and interactivity in their definitions. One possible conclusion about this finding is that better definitions are needed so that everyone has a shared understanding. Or perhaps more language is needed to differentiate between different kinds of embellishments and their functions. Another possible conclusion is that standardized training is needed to ensure conceptual consistency among practitioners. While we do not discount these conclusions altogether, as clear definitions and training are certainly valuable, we argue that a more important takeaway is that we need to understand the unique ways in which designers generate and use knowledge in real-world settings, with a particular focus on the personal and situated nature of design practice.

One interesting finding is that some participants thought of \textit{interaction} as akin to ``junk'' or ``embellishment'', as in something that is superfluous or redundant. While interaction may sometimes be unnecessary and can have costs \cite{lam2008framework}, it can also have significant value for comprehension, especially with complex data \cite{sedig2013interaction}. As we did not explicitly ask our participants about interaction, it is unclear how widespread these views are. However, it is still a topic that could benefit from more investigation. While some recent work has looked at whether adding interaction to static visualizations is beneficial \cite{mosca_ottley_chang_2020}, better design guidance would likely be valuable. 

\subsection{Need for Tools to Support Embellishment Creation}
Multiple participants described embellishments as irrelevant for them, not because they had particular opinions about their use, but because they thought of themselves as not being able to produce embellishments effectively. This finding surfaces questions about the ways in which tools can and should support the activities of visualization designers. For instance, it is unclear whether embellishments can be supported by tools in any kind of standardized fashion, or whether they need to be custom-made in tools like Adobe Illustrator. This may be relevant for discussions that have been taking place in the visualization community regarding challenges in evaluating visualization authoring systems \cite{ren2018reflecting,satyanarayan2019critical} and efforts to study tool use and other practice-oriented issues \cite{bigelow2016iterating,hoffswell2020techniques,mendez2017bottom,parsons_what_2020}. This finding also surfaces questions about the ways in which practitioners view their own design competence and the skills required to be a good designer. While some kinds of embellishments may require artistic skill, others certainly do not. However, without better guidance regarding the use of embellishments, some practitioners may simply see their use as an art that is unattainable for them.

\subsection{Use of Embellishment is Personal and Situated}
Our work shows that the ways in which practitioners use chartjunk in their practice are highly varied and pluralistic, being strongly influenced by personal characteristics and goals of designers, their underlying philosophies, and the situated and pragmatic aspects of the design setting. Although findings from studies on chartjunk suggest when and how its use is appropriate (e.g., to increase memorability or engagement), these recommendations do not fully characterize the considerations that practitioners give towards its use. Many participants were at least vaguely aware of studies on memorability and comprehension, and claimed to be influenced by their findings, although usually not in ways that could be clearly articulated. In addition to any influence these findings had on practitioners, there were clearly other influences that were highly personal and situated. These included descriptions of personal style and preference, skill (or lack thereof) in creating embellishments, constraints from clients and branding, and underlying philosophical commitments about the value and purpose of visualization. 

It is difficult to determine how influential research findings are for these practitioners, especially because people have a tendency in retrospective analysis to make situated activity seem more ``rational'' than it really was \cite{Suchman_1987}. When asked about the ongoing discourse on chartjunk in academic and practitioner spaces, our participants were more likely to be familiar with discussions happening on social media or in practitioner publications than in academic papers and conferences. Thus it is hard to know where the drivers of attitude change lie---e.g., the ``corrective'' movement discussed previously---and whether the trends within practitioner spaces are more influential than findings coming from researchers. While multiple participants described a useful ``correction'' away from extreme minimalism taking place, only P19, who has a PhD in visualization, referred to any specific work in the academic literature.

InfoVis research has valued abstract and rationalistic ways of discussing knowledge and practice, which is typical of disciplines with positivist foundations \cite{meyer_criteria_2020}. As a result, much research aims to remove or ignore the messy, personal, situated aspects of artifacts and their use, which are the exact things that need to be considered if we are to understand visualization design practice \cite{goodman_understanding_2011,Gray_judgment_2015,Stolterman2008}. Scholarship on design practice in other fields indicates that designers tend not to use knowledge that is too prescriptive; rather, designers appreciate individual concepts or high-level ideas that can inspire, can be used metaphorically as thinking tools, and are open to interpretation and pluralistic means of use \cite{bertelsen_design_2000,Stolterman2008}. Designers tend to appropriate concepts and methods, understanding them in their own terms and using them in ways that fit the situation, which may be different from the intentions of their creators \cite{gray_reprioritizing_2014}. Thus, our findings---that chartjunk is not well defined, is interpreted quite broadly, and is influenced by personal and situated concerns--are in line with findings in other design disciplines. It is plausible that the broad interpretations of chartjunk may account, at least partially, for its popularity and continued attention over many decades.

\section{Summary and Future Work}
Our work has shown that chartjunk is understood very broadly and used in a variety of ways by DataVis practitioners. Furthermore, we have shown that the interpretation and use of chartjunk goes beyond simply applying guidelines and research findings in practice settings. Rather, practitioners rely on a host of personal and situated matters when deciding on when and how to use embellishments.

Our work points to a growing movement beyond cognitivism, highlighting a turn to experience as an important framing for discussing and participating in design practice. This movement seems to be a growing in the research community as well (e.g., ~\cite{pousman_casual_2007,saket_beyond_2016,hung_assessing_2017,wang2019emotional}), and perhaps it opens up new ways for researchers to investigate the effects of embellishments on users.  

We hope our investigation surfaces a need for more practice-led research within the visualization community. If InfoVis can follow the lead of other design disciplines in embracing practice-led accounts of design, a more holistic account of how and why concepts such as chartjunk are used in real-world settings can be developed.


\bibliographystyle{abbrv-doi}

\bibliography{template}
\end{document}